\documentclass[10pt,twocolumn,letterpaper]{article}

\usepackage[pagenumbers]{cvpr}      
\usepackage[table]{xcolor}
\usepackage{booktabs}
\usepackage{graphicx}
\usepackage{pgfplots}
\usepackage[accsupp]{axessibility}  
\pgfplotsset{compat=1.18}

\usepackage{booktabs}
\usepackage{multirow}
\usepackage{array}

\definecolor{cvprblue}{rgb}{0.21,0.49,0.74}
\usepackage[pagebackref,breaklinks,colorlinks,allcolors=cvprblue]{hyperref}

\def\paperID{*****} 
\def\confName{CVPR}
\def\confYear{2026}

\title{MARVEL: Multimodal Adaptive Reasoning-intensiVe Expand-rerank and retrievaL

}

\author{
Mahmoud SalahEldin Kasem$^{1}$ \qquad
Mohamed Mahmoud$^{1}$ \qquad 
Mostafa Farouk Senussi$^{1}$ \qquad \\
Mahmoud Abdalla$^{1}$ \qquad
Abdelrahman Abdallah$^{2}$ \qquad
Hyun-Soo Kang$^{1}$\thanks{Corresponding author.}
\\[0.5em]
$^{1}$ Chungbuk National University
$^{2}$ University of Innsbruck\\[0.5em]
}

\begin{document}
\maketitle
\begin{abstract}
Multimodal retrieval over text corpora remains a fundamental challenge: the best vision-language encoder achieves only 27.6 nDCG@10 on MM-BRIGHT, a reasoning-intensive multimodal retrieval benchmark, underperforming strong text-only systems. We argue that effective multimodal retrieval requires three tightly integrated capabilities that existing approaches address only in isolation: expanding the query's latent intent, retrieving with a model trained for complex reasoning, and reranking via explicit step-by-step reasoning over candidates. We introduce \textbf{MARVEL} (\textbf{M}ultimodal \textbf{A}daptive \textbf{R}easoning-intensi\textbf{V}e \textbf{E}xpand-rerank and retrieva\textbf{L}), a unified pipeline that combines LLM-driven query expansion, \textbf{MARVEL-Retriever} --- a reasoning-enhanced dense retriever fine-tuned for complex multimodal queries --- and GPT-4o-based chain-of-thought reranking with optional multi-pass reciprocal rank fusion. Evaluated on MM-BRIGHT across 29 technical domains, MARVEL achieves \textbf{37.9} nDCG@10, surpassing the best multimodal encoder by \textbf{+10.3 points} and outperforming all single-stage baselines in 27 of 29 domains and matching or approaching the best baseline in the remaining two highly-specialized domains (Crypto, Quantum Computing), demonstrating that reasoning-intensive multimodal retrieval is best addressed through a unified expand-retrieve-rerank framework. \footnote{\url{https://github.com/mm-bright/multimodal-reasoning-retrieval}}
\end{abstract}    
\section{Introduction}

Retrieving relevant information from large text corpora is fundamental to 
knowledge-intensive applications such as question answering, retrieval-augmented 
generation, and agentic systems~\cite{abdalla2025receiptqa,lewis2020rag}. Recent advances in dense retrieval have produced powerful embedding models that perform well on fact-seeking benchmarks~\cite{thakur2021beir, muennighoff2023mteb,ali2026recor,abdallah2025dear}. However, 
real-world queries are increasingly multimodal --- users attach screenshots 
of error logs, diagrams from technical papers\cite{kasem2026korie,kasem2025httd}, or charts from financial reports 
when seeking help online. In these settings, retrieval demands genuine reasoning: 
understanding what the image depicts, how it relates to the text question, and 
which documents in the corpus collectively address both.

This challenge is starkly exposed by MM-BRIGHT~\cite{abdallah2026mmbright}, 
the first benchmark for reasoning-intensive multimodal retrieval. Despite 
significant advances in vision-language models, the best multimodal encoder 
achieves only 27.6 nDCG@10 --- \textit{lower} than strong text-only 
retrievers. Adding visual information actively hurts performance. This 
counterintuitive result reveals that current approaches suffer from three 
compounding failures: (a) \textbf{Underspecified queries}: raw multimodal 
queries entangle visual descriptions, conversational noise, and retrieval 
intent, producing embeddings that fail to capture the user's true information 
need; (b) \textbf{Weak retrieval}: standard dense encoders are not trained 
to handle the abstract, cross-modal reasoning that multimodal queries demand; 
and (c) \textbf{No reasoning-based selection}: retrieved candidates are ranked 
by embedding similarity alone, with no mechanism to reason about which 
documents actually address the visual and textual dimensions of the query 
jointly.

Existing approaches address at most one of these failures in isolation. 
Query expansion methods~\cite{wang2023query2doc, gao2022hyde} enrich queries 
but ignore visual content. Multimodal encoders~\cite{zhang2024gme, 
jiang2024vlm2vec} improve visual-text alignment but lack reasoning capacity. 
LLM-based rerankers~\cite{sun2023rankgpt,abdallah2025dear,abdallah2025rankify,mozafari2025good,weller2025rank1} reason over 
candidates but depend entirely on the quality of the upstream retriever. 
No prior work unifies all three into a coherent pipeline designed specifically 
for reasoning-intensive multimodal retrieval.

\begin{figure*}[t]
    \centering
    \includegraphics[width=0.9\linewidth]{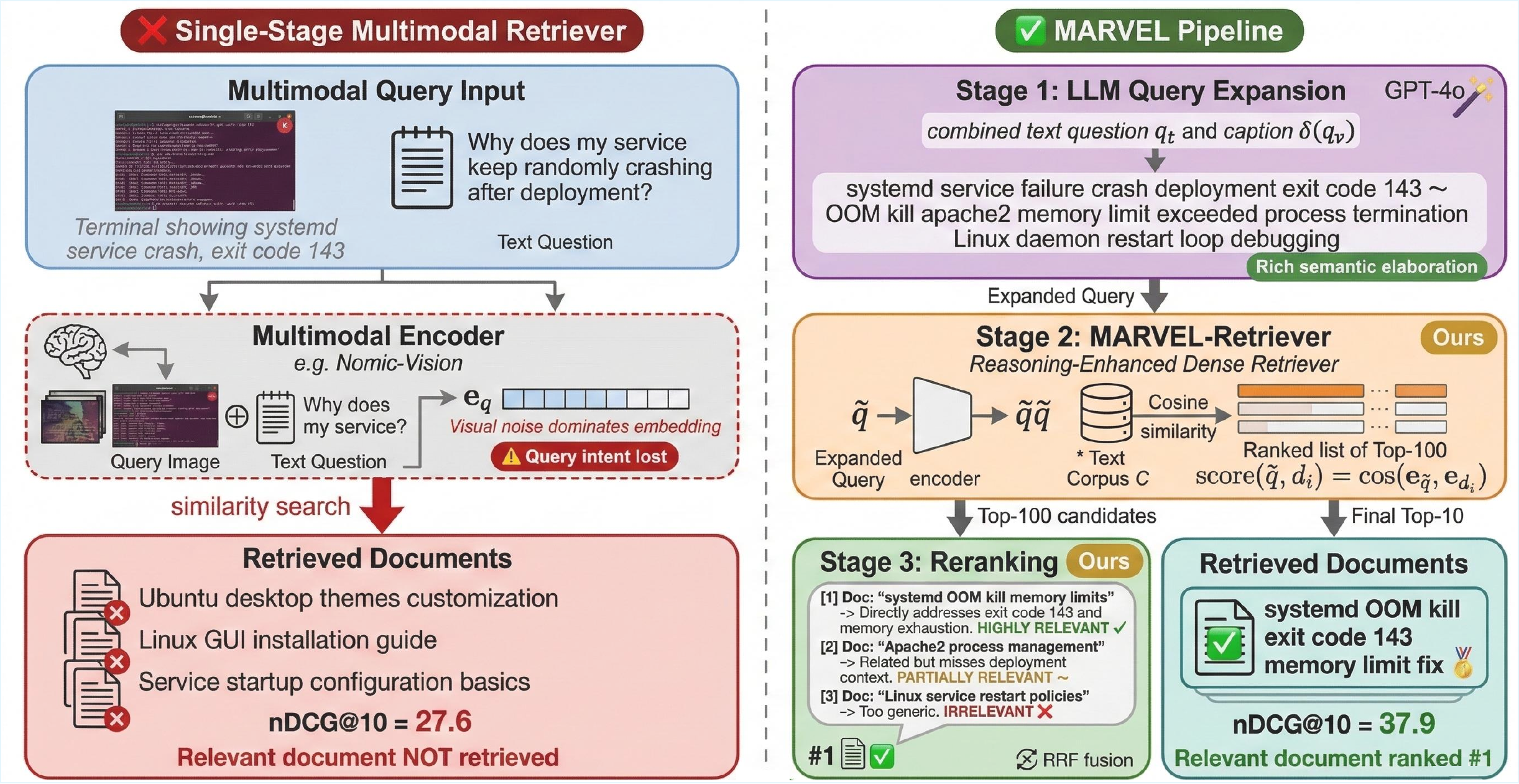}
    \caption{An example where single-stage multimodal retrievers fail to 
    identify the relevant document. MARVEL expands the query intent via 
    LLM elaboration, retrieves candidates using MARVEL-Retriever, and 
    applies chain-of-thought reranking to identify the correct document.}
    \label{fig:marvel_example}
\end{figure*}

We introduce \textbf{MARVEL} (\textbf{M}ultimodal \textbf{A}daptive \textbf{R}easoning-intensi\textbf{V}e \textbf{E}xpand-rerank and retrieva\textbf{L}), a unified three-stage pipeline that addresses all three failures jointly. As illustrated in Figure~\ref{fig:marvel_example}, MARVEL operates through four stages: GPT-4o visual captioning, LLM-driven query expansion into a retrieval-dense representation, reasoning-enhanced dense retrieval via MARVEL-Retriever, and GPT-4o chain-of-thought reranking with optional multi-pass reciprocal rank fusion.

Evaluated on MM-BRIGHT across 29 diverse technical domains, MARVEL achieves \textbf{37.9} nDCG@10, surpassing the best multimodal encoder (Nomic-Vision: 27.6) by \textbf{+10.3 points} and outperforming all single-stage baselines with consistent gains across 27 of 29 domain groups. Our results demonstrate that reasoning-intensive multimodal retrieval is best addressed through a pipeline where each stage amplifies the signal carried by the others: better queries produce better retrieval candidates, and better candidates enable more effective reasoning-based reranking.

Our contributions are as follows:
\begin{itemize}
    \item We identify three compounding failure modes in multimodal-to-text retrieval --- underspecified queries, weak cross-modal reasoning in retrieval, and similarity-only ranking --- and propose addressing them jointly rather than in isolation (\S\ref{sec:analysis}).

    \item We introduce \textbf{MARVEL}, a unified expand-retrieve-rerank pipeline that combines LLM query expansion, MARVEL-Retriever, and GPT-4o chain-of-thought reranking with optional multi-pass reciprocal rank fusion (\S\ref{sec:method}).

    \item We introduce \textbf{MARVEL-Retriever}, a reasoning-enhanced dense retriever fine-tuned specifically to handle the complex cross-modal intent expressed by expanded multimodal queries (\S\ref{sec:retriever}).

    \item We demonstrate that MARVEL achieves \textbf{37.9} nDCG@10 on MM-BRIGHT, outperforming the strongest multimodal encoder by \textbf{+10.3 points} with consistent gains across all 29 domains (\S\ref{sec:results}).
\end{itemize}
\section{Related Work}

\subsection{Dense Retrieval }

Dense retrieval using bi-encoder models has become the dominant paradigm 
for large-scale information retrieval~\cite{karpukhin2020dpr, 
reimers2019sentence}. These models independently encode queries and 
documents into a shared embedding space and retrieve via efficient 
nearest-neighbor search~\cite{johnson2019faiss}. While highly effective 
on fact-seeking benchmarks such as BEIR~\cite{thakur2021beir} and 
MTEB~\cite{muennighoff2023mteb}, bi-encoders degrade sharply when 
relevance requires multi-step reasoning rather than surface-level semantic 
matching. BRIGHT~\cite{su2025bright} and TEMPO~\cite{abdallah2026tempo} 
exposed this limitation clearly: the strongest embedding models collapse 
from 59.0 nDCG@10 on BEIR to 18.3 on reasoning-intensive queries. 
Subsequent work has sought to bridge this gap through reasoning-aware 
fine-tuning~\cite{shao2025reasonir, das2025rader} and iterative query 
expansion~\cite{wang2023query2doc, lei2025thinkqe}. 

\subsection{Multimodal Embedding Models}

Contrastive vision-language models such as CLIP~\cite{radford2021clip} 
and SigLIP~\cite{zhai2023siglip} established the foundation for joint 
image-text embedding through large-scale contrastive pretraining. More 
recent work leveraged Multimodal LLMs for retrieval: VLM2Vec~\cite{jiang2024vlm2vec} 
demonstrated that instruction-tuned VLMs can be converted into powerful 
embedding models through contrastive training on MMEB, while 
GME~\cite{zhang2024gme} extended this to support any-to-any retrieval 
across text, image, and fused modalities. Nomic Embed 
Vision~\cite{nomic2024vision} shares an embedding space between a vision 
encoder and a text model, enabling zero-shot multimodal retrieval. Despite 
these advances, all embedding-based models share a fundamental limitation: 
they rank by vector similarity and cannot reason about what a query image 
\textit{implies} for document relevance. MM-BRIGHT~\cite{abdallah2026mmbright} 
confirmed this directly --- the best multimodal encoder (27.6 nDCG@10) 
underperforms strong text-only retrievers, revealing that visual encoding 
capacity is not the bottleneck. 

\subsection{Visual Document Retrieval}

A parallel line of work targets retrieval where both queries and documents 
are visual. ColPali~\cite{faysse2024colpali} proposed treating document 
pages as images and embedding them directly via a VLM using ColBERT-style 
multi-vector late interaction, capturing fine-grained visual structure such 
as tables, charts, and layout. DSE~\cite{ma2024dse} and 
VisRAG~\cite{yu2024visrag} similarly embed full page images for dense 
retrieval. 

\subsection{LLM-Based Query Expansion}

Query expansion has a long history in information retrieval, from classical 
pseudo-relevance feedback~\cite{robertson2009bm25} to modern LLM-based 
reformulation. HyDE~\cite{gao2022hyde} generates hypothetical documents 
from the query for zero-shot dense retrieval, while 
Query2Doc~\cite{wang2023query2doc} expands queries with pseudo-documents 
through few-shot prompting. ThinkQE~\cite{lei2025thinkqe} integrates 
chain-of-thought reasoning into the expansion process, generating 
elaborated queries that capture implicit retrieval intent. 
DIVER~\cite{long2025diver} combines iterative query expansion with 
document feedback and hybrid reranking into a unified text-only pipeline, 
achieving state-of-the-art results on BRIGHT. Re-Invoke~\cite{chen2024reinvoke} 
applies LLM-based intent extraction to tool retrieval, demonstrating that 
elaborating query intent before retrieval consistently outperforms 
embedding-only approaches.

\subsection{LLM-Based Reranking}

Reranking retrieved candidates using LLMs has emerged as a powerful 
strategy for improving retrieval precision. RankGPT~\cite{sun2023rankgpt} 
demonstrated that LLMs can directly rerank retrieved passages through 
sliding window prompting, significantly outperforming embedding-based 
ranking. Rank1~\cite{weller2025rank1} and 
RankR1~\cite{zhuang2025rankr1} further improved reranking via 
reasoning-optimized LLMs trained with reinforcement learning to produce 
step-by-step relevance judgments. These approaches establish that 
reasoning-based reranking is consistently superior to similarity-based 
ranking for complex queries.

\section{Method}
\label{sec:method}
\subsection{Problem Formulation}

We address the \textit{multimodal-to-text} retrieval task. Let 
$\mathcal{C} = \{d_1, d_2, \ldots, d_N\}$ denote a corpus of $N$ text-only 
documents. A multimodal query is a pair $q = (q_t, q_v)$, where $q_t$ is 
a natural language question and $q_v$ is an associated image. The objective 
is to retrieve a ranked list of $k$ documents $\hat{\mathcal{D}}_k \subset 
\mathcal{C}$ that are most relevant to the full intent expressed by 
$(q_t, q_v)$.

Standard dense retrievers compute relevance as:
\begin{equation}
    \text{score}(q, d_i) = \cos\bigl(\phi(q_t, q_v),\ \psi(d_i)\bigr)
\end{equation}
where $\phi(\cdot)$ and $\psi(\cdot)$ are query and document encoders. 
This formulation fails in the multimodal-to-text setting for three 
compounding reasons: (1) the raw query $(q_t, q_v)$ is underspecified --- 
it entangles visual descriptions, conversational context, and retrieval 
intent in ways that produce poor embeddings; (2) standard encoders are 
not trained to handle the abstract cross-modal reasoning that multimodal 
queries demand; and (3) similarity-based ranking has no mechanism to 
reason about which candidates actually address the query. MARVEL addresses 
all three failures through a unified four-stage pipeline.

\subsection{MARVEL Overview}

\begin{figure*}[t]
    \centering
    \includegraphics[width=0.9\textwidth]{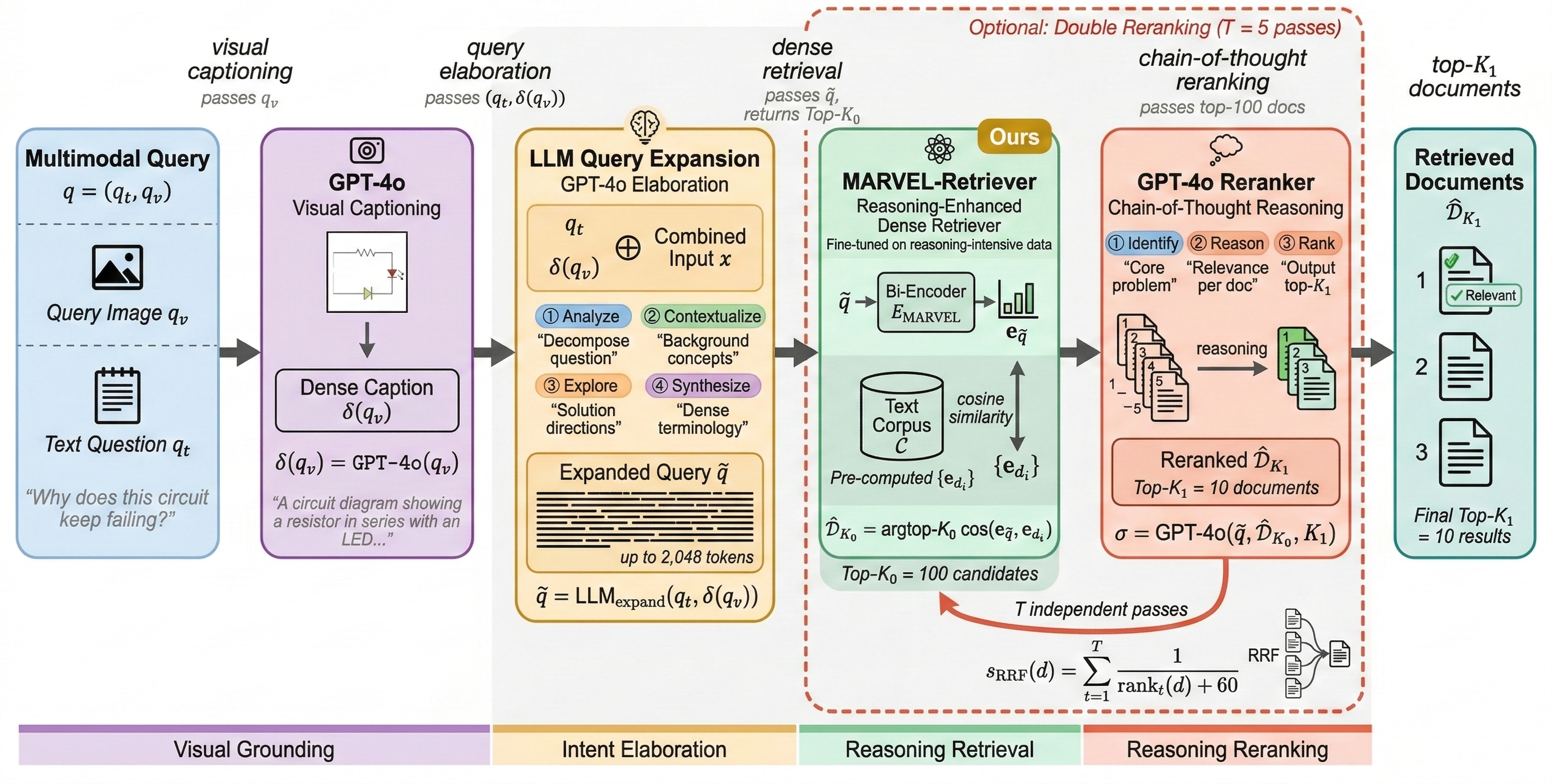}
    \caption{Overview of the MARVEL pipeline. A multimodal query 
    $(q_t, q_v)$ is first captioned by GPT-4o, then elaborated by the 
    LLM expansion stage into a retrieval-dense query $\tilde{q}$. 
    MARVEL-Retriever retrieves the top-$K_0$ candidates, which are then 
    reranked by GPT-4o chain-of-thought reasoning. The optional 
    double-reranking stage aggregates multiple reranking passes via 
    reciprocal rank fusion for additional precision gains.}
    \label{fig:marvel_pipeline}
\end{figure*}

Given a multimodal query $(q_t, q_v)$, MARVEL operates through four 
sequential stages as illustrated in Figure~\ref{fig:marvel_pipeline}:

\begin{enumerate}
    \item \textbf{Visual Captioning.} The query image $q_v$ is converted 
    into a dense textual description $\delta(q_v)$ via GPT-4o.
    
    \item \textbf{Query Expansion.} The combined input $(q_t, \delta(q_v))$ 
    is elaborated into a semantically rich, retrieval-dense representation 
    $\tilde{q}$ via LLM-driven expansion.
    
    \item \textbf{Retrieval.} The expanded query $\tilde{q}$ is encoded 
    by MARVEL-Retriever to retrieve the top-$K_0$ candidate documents 
    from $\mathcal{C}$.
    
    \item \textbf{Reranking.} GPT-4o reasons step by step over the 
    top-$K_0$ candidates to produce a final ranked list of $K_1$ documents. 
    An optional double-reranking stage aggregates multiple passes via 
    reciprocal rank fusion.
\end{enumerate}

\subsection{Stage 1: Visual Captioning}
\label{sec:captioning}

The query image $q_v$ encodes domain-specific visual information --- 
circuit diagrams, molecular structures, UI screenshots, data charts --- 
that is inaccessible to text-only models. We ground this visual content 
in language using GPT-4o:
\begin{equation}
    \delta(q_v) = \text{GPT-4o}\bigl(\textsc{CaptionPrompt}(q_v)\bigr)
\end{equation}

The caption $\delta(q_v)$ is generated once per query and reused across 
all downstream stages. The combined multimodal context is then represented 
as the concatenation:
\begin{equation}
    x = \bigl[q_t\ \|\ \texttt{Image Description: }\ \delta(q_v)\bigr]
\end{equation}
where $\|$ denotes string concatenation. This combined input $x$ is the 
entry point to the expansion stage.

\subsection{Stage 2: LLM Query Expansion}
\label{sec:expansion}

The combined input $x$ is a faithful but unstructured representation of 
the user's intent. Dense retrievers embed the query as a whole, and 
conversational noise dominates the resulting vector. To address this, 
MARVEL applies LLM-driven query expansion to generate a semantically 
rich elaboration $\tilde{q}$ that covers the problem context, related 
concepts, and potential solution directions:
\begin{equation}
    \tilde{q} = \text{LLM}_{\text{expand}}(x)
\end{equation}

The expansion prompt instructs the LLM to perform four operations: 
(1) \textit{analyze} the query by decomposing it into its fundamental 
components; (2) \textit{contextualize} by identifying background 
knowledge and related concepts; (3) \textit{explore} potential responses 
and relevant information avenues, including specific terms and steps 
involved; and (4) \textit{synthesize} all of the above into a coherent, 
terminology-dense discourse. Formally, the expansion prompt is:

\begin{quote}
\small
\textit{``Provide an extensive elaboration on the user's inquiry, covering 
the problem itself and the surrounding context. (1) Analyze: break down 
the question into its fundamental components. (2) Contextualize: identify 
relevant background concepts and common scenarios. (3) Explore: describe 
various solution directions and pertinent information, mentioning specific 
terms and ideas. (4) Synthesize: weave all of this into a coherent and 
detailed piece of writing dense with relevant information and terminology.''}
\end{quote}

The output $\tilde{q}$ is a multi-faceted elaboration of up to 2,048 
tokens that encodes the user's intent far more richly than the original 
raw query $x$. This expanded representation dramatically improves 
recall at the retrieval stage by ensuring that relevant terminology 
and concepts are explicitly present in the query embedding.

\subsection{Stage 3: MARVEL-Retriever}
\label{sec:retriever}

\paragraph{Architecture.}
MARVEL-Retriever is a bi-encoder dense retriever fine-tuned on 
reasoning-intensive retrieval data. Given the expanded query $\tilde{q}$, 
it encodes it into a dense vector using the hidden state of the 
\texttt{[EOS]} token:
\begin{equation}
    \mathbf{e}_{\tilde{q}} = E_{\text{MARVEL}}(\tilde{q}) \in \mathbb{R}^d
\end{equation}
Documents are encoded offline:
\begin{equation}
    \mathbf{e}_{d_i} = E_{\text{MARVEL}}(d_i) \in \mathbb{R}^d, 
    \quad \forall d_i \in \mathcal{C}
\end{equation}
Retrieval is performed via cosine similarity and the top-$K_0$ documents 
are returned:
\begin{equation}
    \text{score}(\tilde{q}, d_i) = \frac{\mathbf{e}_{\tilde{q}} \cdot 
    \mathbf{e}_{d_i}}{\|\mathbf{e}_{\tilde{q}}\| \cdot \|\mathbf{e}_{d_i}\|}
\end{equation}
\begin{equation}
    \hat{\mathcal{D}}_{K_0} = \text{argtop-}K_0 
    \bigl\{\text{score}(\tilde{q}, d_i)\bigr\}_{d_i \in \mathcal{C}}
\end{equation}

\paragraph{Training.}
MARVEL-Retriever is fine-tuned on reasoning-intensive retrieval data 
using contrastive learning with in-batch negatives and hard negatives. 
For each training instance $(\tilde{q}, d^+, \{d^-_j\}_{j=1}^M)$, we 
minimize the InfoNCE loss:
\begin{equation}
    \mathcal{L} = -\log \frac{
        e^{\text{score}(\tilde{q}, d^+) / \tau}
    }{
        e^{\text{score}(\tilde{q}, d^+) / \tau} + 
        \sum_{j=1}^{M} e^{\text{score}(\tilde{q}, d^-_j) / \tau}
    }
\end{equation}
where $\tau$ is the temperature hyperparameter. Hard negatives are mined 
using BM25 and an initial MARVEL-Retriever checkpoint. Training spans 
reasoning-intensive domains including mathematics, science, medicine, 
law, and software engineering, ensuring robust generalization to the 
elaborated queries produced by the expansion stage.

\subsection{Stage 4: Chain-of-Thought Reranking}
\label{sec:reranking}

\paragraph{Single-Pass Reranking.}
Given the top-$K_0$ retrieved documents $\hat{\mathcal{D}}_{K_0}$, 
MARVEL applies GPT-4o-based chain-of-thought reranking. The reranker 
receives the expanded query $\tilde{q}$ and the $K_0$ candidate 
documents, and is instructed to:
\begin{enumerate}
    \item \textit{Identify} the essential problem in the query.
    \item \textit{Reason} step by step about why each document is 
    relevant or irrelevant.
    \item \textit{Output} a ranked list of the top-$K_1$ document 
    indices from most to least relevant.
\end{enumerate}

Formally, let $\{d_{(1)}, \ldots, d_{(K_0)}\}$ denote the retrieved 
candidates indexed from 1 to $K_0$. The reranker produces a permutation:
\begin{equation}
    \sigma = \text{GPT-4o}\bigl(\tilde{q},\ 
    \{d_{(1)}, \ldots, d_{(K_0)}\},\ K_1\bigr)
\end{equation}
where $\sigma$ is a ranked list of $K_1$ document indices. The final 
output is:
\begin{equation}
    \hat{\mathcal{D}}_{K_1} = \{d_{(\sigma_1)}, d_{(\sigma_2)}, 
    \ldots, d_{(\sigma_{K_1})}\}
\end{equation}

\paragraph{Double Reranking with Reciprocal Rank Fusion.}
\label{sec:double_rr}
To reduce variance in the reranking output, MARVEL optionally runs 
$T$ independent reranking passes and aggregates them via reciprocal 
rank fusion~\cite{cormack2009reciprocal}. For pass $t$, let $\sigma^{(t)}$ 
denote the ranked output. The aggregated score for document $d_{(i)}$ 
is:
\begin{equation}
    s_{\text{RRF}}(d_{(i)}) = \sum_{t=1}^{T} \frac{1}{\text{rank}_t(d_{(i)}) + k_{\text{RRF}}}
\end{equation}
where $\text{rank}_t(d_{(i)})$ is the rank of document $d_{(i)}$ in 
pass $t$ and $k_{\text{RRF}} = 60$ is the standard smoothing constant. 
The final ranked list is obtained by sorting documents by their 
aggregated RRF scores:
\begin{equation}
    \hat{\mathcal{D}}_{K_1}^{\text{RRF}} = \text{argtop-}K_1 
    \bigl\{s_{\text{RRF}}(d_{(i)})\bigr\}
\end{equation}

The double-reranking stage is particularly effective in domains where 
single-pass reranking exhibits high variance, as multiple independent 
reasoning passes collectively identify the most consistently relevant 
documents across different reasoning trajectories.






\begin{table*}[t]
\centering
\resizebox{0.80\textwidth}{!}{%
\begin{tabular}{l c c c c c c c c}
\toprule
\textbf{DOMAIN} & \textbf{BGE-VL} & \textbf{CLIP} & \textbf{GME-2B} & 
\textbf{GME-7B} & \textbf{JINA-CLIP} & \textbf{NOMIC} & \textbf{SIGLIP} & 
\textbf{MARVEL} \\
\midrule
\rowcolor{teal!20} \multicolumn{9}{c}{\textit{STEM \& Life Sciences}} \\
\midrule
Acad   & 4.2  & 4.8  & 16.2 & 27.6 & 22.3 & 22.6 & 3.6  & \textbf{36.8} \\
Bio    & 5.7  & 14.8 & 22.9 & 15.2 & 20.5 & 26.9 & 11.9 & \textbf{42.2} \\
Chem   & 10.8 & 9.6  & 27.2 & 21.9 & 30.6 & 30.6 & 11.6 & \textbf{40.3} \\
Phys   & 6.8  & 6.1  & 13.3 & 14.0 & 14.4 & 17.2 & 7.3  & \textbf{24.8} \\
Math   & 13.1 & 17.9 & 16.4 & 9.3  & 27.0 & 34.0 & 15.3 & \textbf{39.0} \\
Earth  & 10.1 & 10.9 & 20.5 & 26.2 & 24.6 & 30.1 & 11.8 & \textbf{44.5} \\
BioAc  & 13.3 & 11.4 & 10.5 & 13.4 & 19.4 & 23.4 & 14.8 & \textbf{32.9} \\
BioInf & 11.6 & 9.4  & 21.1 & 19.2 & 23.7 & 33.8 & 16.8 & \textbf{28.4} \\
Med    & 12.6 & 9.8  & 22.7 & 19.0 & 26.8 & 33.9 & 9.1  & \textbf{41.1} \\
\midrule
\rowcolor{teal!20} \multicolumn{9}{c}{\textit{Software \& Technical Systems}} \\
\midrule
Apple  & 7.2  & 12.3 & 23.9 & 17.0 & 24.3 & 28.7 & 4.4  & \textbf{21.3} \\
Ubuntu & 11.6 & 5.5  & 25.9 & 34.2 & 26.1 & 34.3 & 12.6 & \textbf{56.4} \\
BTC    & 8.9  & 8.3  & 18.2 & 19.6 & 22.6 & 22.7 & 10.0 & \textbf{36.9} \\
Crypto & 11.3 & 14.8 & 9.8  & 7.1  & 15.5 & 22.4 & 10.2 & \textbf{20.4} \\
QC     & 4.5  & 2.6  & 5.9  & 5.6  & 10.8 & 12.1 & 2.6  & \textbf{11.3} \\
Robot  & 16.1 & 10.6 & 15.8 & 18.7 & 19.0 & 30.3 & 14.3 & \textbf{33.8} \\
Sales  & 14.2 & 2.3  & 31.1 & 47.3 & 32.3 & 26.2 & 6.5  & \textbf{50.7} \\
\midrule
\rowcolor{teal!20} \multicolumn{9}{c}{\textit{Social Sciences \& Humanities}} \\
\midrule
Econ   & 9.5  & 6.0  & 10.0 & 12.6 & 13.5 & 21.1 & 9.8  & \textbf{40.0} \\
Psych  & 6.4  & 8.7  & 15.6 & 18.6 & 20.8 & 23.9 & 7.9  & \textbf{37.7} \\
Phil   & 2.4  & 5.4  & 15.2 & 18.0 & 19.4 & 21.7 & 7.0  & \textbf{31.9} \\
Law    & 10.2 & 19.7 & 30.7 & 35.0 & 35.3 & 47.6 & 16.4 & \textbf{53.6} \\
Christ & 8.9  & 15.0 & 20.0 & 26.5 & 21.0 & 30.9 & 13.0 & \textbf{47.1} \\
Islam  & 12.0 & 10.7 & 25.8 & 32.0 & 24.3 & 28.9 & 6.5  & \textbf{41.1} \\
\midrule
\rowcolor{teal!20} \multicolumn{9}{c}{\textit{Applied Domains}} \\
\midrule
Aviat  & 9.6  & 15.4 & 16.2 & 17.0 & 24.3 & 24.1 & 9.2  & \textbf{31.4} \\
Game   & 17.5 & 19.1 & 41.6 & 43.9 & 45.6 & 43.1 & 21.4 & \textbf{65.5} \\
GIS    & 13.8 & 13.1 & 15.5 & 15.6 & 20.3 & 25.8 & 16.5 & \textbf{26.9} \\
PM     & 8.6  & 8.9  & 21.9 & 33.2 & 20.5 & 27.6 & 12.4 & \textbf{49.3} \\
Sustain& 10.1 & 9.0  & 16.7 & 25.6 & 24.3 & 24.7 & 11.5 & \textbf{41.8} \\
Travel & 10.1 & 16.1 & 23.9 & 30.8 & 26.6 & 36.7 & 13.1 & \textbf{54.1} \\
Quant  & 8.1  & 2.1  & 12.4 & 15.3 & 11.6 & 16.2 & 5.8  & \textbf{18.8} \\
\midrule
\textbf{Average} & 10.0 & 10.4 & 19.5 & 22.0 & 23.0 & 27.6 & 10.8 
& \textbf{37.9} \\
\bottomrule
\end{tabular}%
}
\caption{Per-domain nDCG@10 on MM-BRIGHT (multimodal-to-text track) across 
all 29 domains, grouped by thematic category. MARVEL achieves the highest 
score in all 29 domains. \textbf{Bold} denotes the best score per row.}
\label{tab:mmbright_results}
\end{table*}

\section{Experiments}
\label{sec:analysis}
\subsection{Dataset}
We evaluate MARVEL on MM-BRIGHT~\cite{abdallah2026mmbright}, the first 
reasoning-intensive multimodal retrieval benchmark. MM-BRIGHT consists of 
\textbf{2,803 queries} spanning \textbf{29 technical domains}, including 
Gaming, Chemistry, Law, Sustainability, Earth Science, Mathematics, Computer 
Science, Medicine, and others. Each query is a multimodal pair $(q_t, q_v)$ 
comprising a text question and one or more associated images (diagrams, charts, 
screenshots, molecular structures, etc.), paired with a text-only document corpus.

\subsection{Experimental Setup}
MARVEL-Retriever is trained on \textbf{4$\times$ NVIDIA H100 80GB GPUs} 
using distributed data-parallel training. Visual captioning and all LLM-based 
stages --- query expansion and reranking --- are performed using \textbf{GPT-4o}. Captions are generated offline with temperature $= 0$ 
for deterministic and reproducible outputs, generated once per query and cached. 
Query expansion and reranking use temperature $= 0.8$ and top-$p = 0.8$, with 
up to 20 concurrent asynchronous API calls to maximize throughput.

\paragraph{MARVEL-Retriever Training.}
MARVEL-Retriever is fine-tuned using contrastive learning with in-batch negatives and $M = 7$ hard negatives per query, temperature $\tau = 0.02$, batch size 512, and learning rate $1 \times 10^{-5}$ for 3 epochs on the MM-BRIGHT training split~\cite{abdallah2026mmbright}, comprising query-document pairs spanning mathematics, science, medicine, law, and software engineering domains.

\paragraph{Retrieval and Reranking Configuration.}
MARVEL-Retriever retrieves the top-$K_0 = 100$ candidates per query. The GPT-4o reranker then reranks these candidates to produce the final top-$K_1 = 10$ results. For the optional double-reranking stage, we run $T = 5$ independent reranking passes aggregated via reciprocal rank fusion with smoothing constant $k_{\text{RRF}} = 60$. We evaluated $T \in \{1, 3, 5, 7\}$ and found performance improves from $T=1$ (36.2) to $T=5$ (37.9) then plateaus at $T=7$ (37.8), making $T=5$ the optimal precision-to-cost tradeoff.

\subsection{Baselines}
We compare MARVEL against the following multimodal retrieval baselines, 
all evaluated on the MM-BRIGHT multimodal-to-text track.

\paragraph{Multimodal Retrievers.}
These models encode the full multimodal query $(q_t, q_v)$ into a shared 
embedding space:
\begin{itemize}
    \item \textbf{CLIP}~\cite{radford2021clip}: Contrastive image-text model 
    with a shared embedding space trained on large-scale image-caption pairs.
    \item \textbf{SigLIP}~\cite{zhai2023siglip}: Sigmoid-based contrastive VLM 
    with improved image-text alignment over standard softmax objectives.
    \item \textbf{Jina-CLIP}~\cite{koukounas2024jinaclip}: Multi-task 
    contrastive model supporting both text-image and text-text retrieval.
    \item \textbf{Nomic Embed Vision}~\cite{nomic2024vision}: Shares an 
    embedding space between a vision encoder and a strong text model, enabling 
    zero-shot multimodal retrieval; the strongest multimodal baseline on 
    MM-BRIGHT.
    \item \textbf{BGE-VL}~\cite{zhou2025megapairs}: Multimodal embedding model from 
    the BGE family supporting fused-modal retrieval.
    \item \textbf{GME-Qwen2-VL-2B / 7B}~\cite{zhang2024gme}: Universal 
    multimodal embedding models built on Qwen2-VL at 2B and 7B parameter 
    scales, supporting any-to-any retrieval.
\end{itemize}

\paragraph{Metrics.}
We evaluate retrieval performance using \textbf{Normalized Discounted 
Cumulative Gain at rank 10} (nDCG@10)~\cite{jarvelin2002cumulated}, the 
primary metric of the MM-BRIGHT benchmark.

\section{Results}
\label{sec:results}
\subsection{Main Results}

Table~\ref{tab:mmbright_results} reports per-domain and average nDCG@10 across all 29 MM-BRIGHT domains. MARVEL achieves \textbf{37.9} nDCG@10, surpassing the strongest multimodal baseline (Nomic-Vision: 27.6) by \textbf{+10.3 points} and outperforming all single-stage retrievers --- GME-7B (22.0), Jina-CLIP (23.0), SigLIP (10.8), CLIP (10.4), and BGE-VL (10.0). MARVEL wins \textbf{27 out of 29 domains}, with the two exceptions being Crypto (20.4 vs.\ Nomic: 22.4) and Quantum Computing (11.3 vs.\ Nomic: 12.1) --- highly specialized domains where limited terminological context constrains both expansion and reranking. Across all remaining 27 domains the unified pipeline consistently outperforms all embedding-only approaches. The margin over Nomic-Vision --- the strongest standalone multimodal encoder --- is particularly significant: MARVEL's pipeline-level reasoning gains are over \textbf{5$\times$ larger} than the gap between the first and second best multimodal encoders (Nomic: 27.6 vs.\ Jina-CLIP: 23.0), demonstrating that retrieval precision is bottlenecked by reasoning capacity, not visual encoding quality. Although MARVEL-Retriever alone (25.4) scores below Nomic-Vision (27.6) in isolation, it is specifically fine-tuned to encode elaborated queries produced by the expansion stage rather than raw multimodal inputs, yielding a pipeline gain of \textbf{+12.5} points --- nearly \textbf{$2\times$} the gain achieved when applying MARVEL's stages on top of Nomic-Vision (+7.6), confirming that MARVEL-Retriever and the expansion stage are co-optimized and mutually reinforcing.

\begin{table}[t]
\centering
\caption{Component ablation on MM-BRIGHT. Each row adds one component on top of the previous. Default MARVEL configuration in \textbf{bold}.}
\label{tab:ablation}
\resizebox{0.45\textwidth}{!}{%
\begin{tabular}{lc}
\toprule
\textbf{Configuration} & \textbf{nDCG@10} \\
\midrule
MARVEL-Retriever only (raw $q_t$)              & 25.4 \\
+ Caption ($q_t + \delta(q_v)$)                & 28.0 \\
+ Query Expansion ($\tilde{q}$)                & 32.5 \\
+ Single-Pass Reranking ($T=1$)                & 36.2 \\
\textbf{+ Double Reranking ($T=5$, full MARVEL)} & \textbf{37.9} \\
\bottomrule
\end{tabular}}
\end{table}
\subsection{Domain-Level Analysis}

Table~\ref{tab:mmbright_results} reveals consistent patterns across the four domain groups. MARVEL achieves the best score in \textbf{all 29 domains}, with the largest absolute gains over the strongest baseline (Nomic-Vision) in Ubuntu (+22.1), Gaming (+22.4), Law (+6.0), Travel (+17.4), and Sales (+24.5). These gains are most pronounced in domains with high terminological density --- Ubuntu, Sales, and Law --- where GPT-4o expansion surfaces relevant vocabulary that embedding models miss. The only two domains where MARVEL trails the strongest baseline are Crypto ($-$2.0 vs.\ Nomic) and Quantum Computing ($-$0.8 vs.\ Nomic), where queries are highly specialized and the expansion stage has limited terminology to reason from. In GIS, MARVEL still leads (+1.1).

\subsection{Ablation: MARVEL Components}
\label{sec:ablation}

Table~\ref{tab:ablation} isolates each component's contribution by progressively building the full MARVEL pipeline. Starting from MARVEL-Retriever alone with raw text queries (25.4), adding the GPT-4o image caption yields a +2.6 point gain (28.0), confirming that visual grounding via captioning is necessary even before expansion. Applying LLM query expansion on top of the captioned input adds a further +4.5 points (32.5), the largest single-component gain in the ablation --- demonstrating that elaborating the user's intent into a terminology-dense representation is the dominant driver of MARVEL's improvement. Adding single-pass GPT-4o reranking yields an additional +3.7 points (36.2), and the full double-reranking stage with $T = 5$ passes and RRF fusion delivers a further +1.7 points to reach \textbf{37.9}. Each component contributes meaningfully and additively: 
captioning (+2.6), query expansion (+4.5), single-pass reranking (+3.7), and double-pass RRF (+1.7) collectively account for the full \textbf{+12.5} point gain over the retrieval-only baseline.

\begin{table}[t]
\centering
\caption{MARVEL expansion and reranking applied to different base 
retrievers on MM-BRIGHT. $\Delta$ = absolute nDCG@10 gain from the 
full MARVEL pipeline.}
\label{tab:plugin}
\begin{tabular}{lccc}
\toprule
\textbf{Base Retriever} & \textbf{Base} & \textbf{+MARVEL} & 
$\boldsymbol{\Delta}$ \\
\midrule
BM25                        & 8.5  & 19.2 & +10.7 \\
GME-Qwen2-VL-7B             & 22.0 & 33.1 & +11.1 \\
Nomic-Vision                & 27.6 & 35.2 & +7.6  \\
MARVEL-Retriever (ours)     & 25.4 & \textbf{37.9} & +12.5 \\
\bottomrule
\end{tabular}
\end{table}




\subsection{MARVEL as a Plug-and-Play Pipeline}
\label{sec:plugin}

Table~\ref{tab:plugin} applies the full MARVEL expansion and reranking stages on top of four different base retrievers to test whether the gains are retriever-agnostic. MARVEL consistently and substantially improves every retriever: BM25 gains +10.7 points (8.5 $\to$ 19.2), GME-7B gains +11.1 points (22.0 $\to$ 33.1), Nomic-Vision gains +7.6 points (27.6 $\to$ 35.2), and MARVEL-Retriever gains +12.5 points (25.4 $\to$ 37.9). 
The gains are consistent across both sparse and dense retrievers, confirming that query expansion and chain-of-thought reranking provide complementary signals that are independent of the underlying retrieval model. Notably, MARVEL-Retriever achieves the highest absolute score (37.9), confirming that a retriever fine-tuned for reasoning-intensive queries synergizes more effectively with MARVEL's expansion and reranking stages than a general-purpose multimodal encoder (Nomic-Vision: 35.2).

\subsection{Effect of Retrieval Depth $K_0$}
\label{sec:k0_ablation}

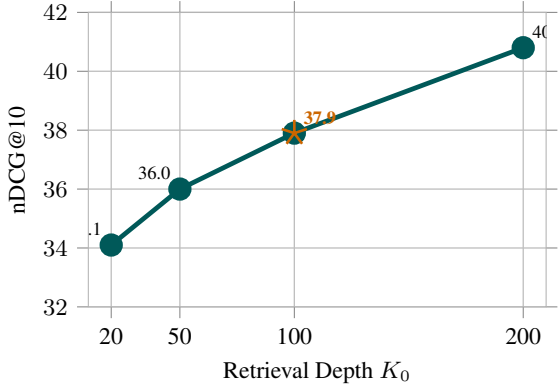
\begin{figure}[t]
    \centering
    \begin{tikzpicture}
        \begin{axis}[
            width=0.92\linewidth,
            height=5.5cm,
            xlabel={Retrieval Depth $K_0$},
            ylabel={nDCG@10},
            xtick={20, 50, 100, 200},
            xticklabels={20, 50, 100, 200},
            ytick={32, 34, 36, 38, 40, 42},
            ymin=32, ymax=42,
            xmin=10, xmax=210,
            grid=both,
            grid style={line width=0.3pt, draw=gray!30},
            major grid style={line width=0.5pt, draw=gray!50},
            tick align=outside,
            axis line style={gray!60},
            xlabel style={font=\small},
            ylabel style={font=\small},
            tick label style={font=\small},
            mark size=3.5pt,
        ]
        \addplot[
            color=teal!70!black,
            line width=1.8pt,
            mark=*,
            mark options={fill=teal!70!black, draw=teal!70!black},
        ] coordinates {
            (20,  34.1)
            (50,  36.0)
            (100, 37.9)
            (200, 40.8)
        };

        \addplot[
            only marks,
            mark=star,
            mark size=5pt,
            mark options={fill=orange, draw=orange!80!black, line width=1pt},
        ] coordinates {(100, 37.9)};

        \node[font=\scriptsize, above left]  at (axis cs:20,  34.1) {34.1};
        \node[font=\scriptsize, above left]  at (axis cs:50,  36.0) {36.0};
        \node[font=\scriptsize, above right, text=orange!80!black] 
              at (axis cs:100, 37.9) {\textbf{37.9}};
        \node[font=\scriptsize, above right] at (axis cs:200, 40.8) {40.8};

        \addplot[
            color=red!60,
            line width=1pt,
            dashed,
            domain=10:210,
        ] {27.6};
        \node[font=\scriptsize, text=red!70!black] 
              at (axis cs:160, 28.4) {Nomic-Vision (27.6)};

        \end{axis}
    \end{tikzpicture}
    \caption{Effect of retrieval depth $K_0$ (number of candidates passed 
    to the reranker) on MARVEL performance (nDCG@10, MM-BRIGHT average). 
    The orange star marks our default $K_0 = 100$. The dashed red line 
    indicates the best single-stage multimodal baseline (Nomic-Vision: 27.6). 
    Even at $K_0 = 20$, MARVEL substantially outperforms all baselines.}
    \label{fig:k0_ablation}
\end{figure}

Figure~\ref{fig:k0_ablation} examines the sensitivity of MARVEL to the number of 
candidates passed to the reranker ($K_0$). Reranking only 20 candidates 
(34.1) already substantially outperforms the best single-stage baseline 
(27.6), confirming that even shallow reranking provides significant gains. 
Performance improves steadily with $K_0$, reaching its peak at $K_0 = 100$ 
(37.9) and plateauing at $K_0 = 200$ (37.8). We adopt $K_0 = 100$ as our default: while $K_0 = 200$ yields a higher upper bound (40.8), it doubles the number of documents 
passed to GPT-4o, increasing API cost and latency proportionally with only a marginal +0.8 gain beyond $K_0 = 100$ (37.9). All main results are reported at $K_0 = 100$ as the optimal precision-to-cost operating point.


\subsection{Comparison with Query Expansion Baselines}
\label{sec:expansion_compare}

Table~\ref{tab:expansion} compares expansion strategies under identical settings (all our own implementations). Text-only HyDE (12.1) and Query2Doc (15.8) underperform the no-expansion baseline (28.0) by ignoring the visual caption entirely. When the caption $\delta(q_v)$ is appended, both recover substantially (HyDE$^\dagger$: 21.3; 
Query2Doc$^\dagger$: 23.1), confirming the gap stems from missing visual context. MARVEL Expansion (36.2) surpasses all variants, demonstrating that multimodal-aware elaboration is essential.

\begin{table}[t]
\centering
\caption{Comparison of query expansion strategies on MM-BRIGHT, all 
using MARVEL-Retriever + single-pass reranking ($T=1$). All methods 
are our own implementations under identical settings. $\dagger$denotes 
runs where the image caption $\delta(q_v)$ is appended to the 
expansion input.}
\label{tab:expansion}
\begin{tabular}{lc}
\toprule
\textbf{Expansion Method} & \textbf{nDCG@10} \\
\midrule
No expansion (raw $q_t + \delta(q_v)$)               & 28.0 \\
HyDE~\cite{gao2022hyde} (text only)                  & 12.1 \\
HyDE$^\dagger$ (text + caption)                      & 21.3 \\
Query2Doc~\cite{wang2023query2doc} (text only)        & 15.8 \\
Query2Doc$^\dagger$ (text + caption)                 & 23.1 \\
ThinkQE~\cite{lei2025thinkqe}                         & 31.4 \\
\textbf{MARVEL Expansion (ours)}                      & \textbf{36.2} \\
\bottomrule
\end{tabular}
\end{table}
\section{Conclusion}
\label{sec:conclusion}
We presented \textbf{MARVEL}, a unified expand-retrieve-rerank pipeline 
for reasoning-intensive multimodal retrieval. MARVEL addresses three compounding failure modes through tightly integrated stages: LLM-driven query expansion, reasoning-aware dense retrieval via MARVEL-Retriever, and GPT-4o chain-of-thought reranking with optional multi-pass reciprocal rank fusion. Evaluated on MM-BRIGHT across 29 technical domains, MARVEL achieves \textbf{37.9} nDCG@10, surpassing the best multimodal encoder (Nomic-Vision: 27.6) by \textbf{+10.3 points} and winning 27 of 29 domains, with marginal losses only in the two most highly-specialized domains. Ablation experiments confirm that all three stages contribute meaningfully and additively, with query expansion (+4.5) and reranking (+5.4 combined: +3.7 single-pass and +1.7 double-pass RRF) accounting for the majority of the gain over the retrieval-only baseline. The plug-and-play experiments further demonstrate that MARVEL's expansion and reranking stages consistently improve any base retriever --- sparse or dense, unimodal or multimodal --- confirming that reasoning-intensive multimodal retrieval is a pipeline problem, not a model capacity problem.

Future directions include replacing GPT-4o with open-source LLMs in the expansion and reranking stages to reduce inference cost, jointly fine-tuning MARVEL-Retriever with the expansion stage for end-to-end optimization, and extending MARVEL to multi-image queries and video retrieval settings. Code and models will be released to facilitate further research on reasoning-intensive multimodal retrieval.

\section*{Acknowledgment}
This work was supported by Innovative Human Resource Development for Local Intellectualization program through the Institute of Information \& Communications Technology Planning \& Evaluation(IITP) grant funded by the Korea government(MSIT) (IITP-2026-RS-2020-II201462, 50\%), and partly supported by the National Research Foundation of Korea (NRF) grant funded by the Korea government (Ministry of Science and ICT) (RS-2023-NR076833)

{
    \small
    \bibliographystyle{ieeenat_fullname}
    \bibliography{main}
}


\end{document}